\documentclass[preprint,nofootinbib]{revtex4-1}
\usepackage{graphicx,epsfig,rotate,hyperref}

\def\iso#1#2{\mbox{${}^{#2}{\rm #1}$}}
\def\he#1{\iso{He}{#1}}
\def\li#1{\iso{Li}{#1}}
\def\be#1{\iso{Be}{#1}}

\def\e10{\eta_{10}}

\def\Ombh2{\Omega_{\rm b} h^2}

\def\ga{\mathrel{\raise.3ex\hbox{$>$\kern-.75em\lower1ex\hbox{$\sim$}}}}
\def\la{\mathrel{\raise.3ex\hbox{$<$\kern-.75em\lower1ex\hbox{$\sim$}}}}

\def\beq{\begin{equation}}
\def\eeq{\end{equation}}
\def\beqar{\begin{eqnarray}}
\def\eeqar{\end{eqnarray}}

\begin{document}

\rightline{UMN--TH--4210/23}
\rightline{FTPI--MINN--23/04}
\rightline{March 2023}

\title{The Neutron Mean Life and Big Bang Nucleosynthesis}\footnote{Invited contribution to the Special Issue of the journal Universe on the Neutron Lifetime, guest editor: B. Grinstein}

\author{Tsung-Han Yeh}
\affiliation{TRIUMF, 4004 Wesbrook Mall, Vancouver, BC V6T 2A3, Canada}
\author{Keith A. Olive}
\affiliation{William I. Fine Theoretical Physics Institute,
School of Physics and Astronomy,
University of Minnesota, Minneapolis, MN 55455, USA}
\author{Brian D. Fields}
\affiliation{Departments of Astronomy and of Physics, and Illinois Center for Advanced Study of the Universe, University of Illinois,
Urbana, IL 61801, USA}

\begin{abstract}
We explore the effect of neutron lifetime and its uncertainty on standard big-bang nucleosynthesis (BBN).  
BBN describes the cosmic production of the light nuclides \iso{H}{1}, D, \iso{H}{3}+\he3, \he4, and \li7+\be7 in the first minutes of cosmic time.  The neutron mean life $\tau_n$ has two roles in modern BBN calculations: (1) it normalizes the matrix element for weak $n \leftrightarrow p$ interconversions, and (2) it sets the rate of free neutron decay after the weak interactions freeze out.  We  review the history of the interplay between $\tau_n$ measurements and BBN, and  present a study of the sensitivity of the light element abundances to the modern neutron lifetime measurements.  We find that $\tau_n$ uncertainties dominate the predicted \he4 error budget, but these theory errors remain smaller than the uncertainties in \he4 observations, even with the dispersion in recent neutron lifetime measurements.  For the other light-element predictions, $\tau_n$ contributes negligibly to their error budget.  Turning the problem around, we combine present BBN and cosmic microwave background (CMB) determinations of the cosmic baryon density to {\em predict} a ``cosmologically preferred'' mean life of $\tau_{n}({\rm BBN+CMB}) = 870 \pm 16 \ \rm sec$, which is consistent with experimental mean life determinations.   We go on to show that if future astronomical and cosmological helium observations can reach an uncertainty of $\sigma_{\rm obs}(Y_p) = 0.001$ in the \he4 mass fraction $Y_p$, this could begin to discriminate between the mean life determinations.
\end{abstract}

\pacs{}
\keywords{}

\maketitle

\section{Introduction}
The origin of the light elements (\iso{H}{1}, D, \he3, \he4, and \li7) is well explained by big bang nucleosynthesis (BBN) taking place in the early Universe when the temperature scale was roughly 1 MeV and below, i.e., the energies associated with nuclear reactions \cite{bbn,iocco,coc18,CFOY,foyy}. 
Because the physics at these energies is well known, testable predictions of BBN are possible given reliable nuclear cross sections and the subject of this contribution, a well-measured neutron lifetime \cite{ParticleDataGroup:2022pth}.  

Standard BBN is the theoretical framework that
implements these cross sections in the context of both the Standard Models of particle physics and cosmology, and is the model we will adopt for this paper. For cosmology, we will thus assume $\Lambda$CDM. For BBN, this means that the Universe was dominated by photon and neutrino radiation during BBN, with a baryon density consistent with that determined from measurements of the cosmic microwave background (CMB) anisotropies \cite{Planck2018}; it is useful to parameterize the baryon density in terms of baryon-to-photon ratio $\eta \equiv n_{\rm b}/n_\gamma$, with the density of blackbody photons fixed by the temperature: $n_\gamma \propto T^3$. From the Standard model of electroweak interactions, we will assume that the number of light neutrino flavors is $N_\nu = 3$ and that there are no other relativistic contributions to the energy density. 

\begin{figure}[!htb]
\includegraphics[width=0.80\textwidth]{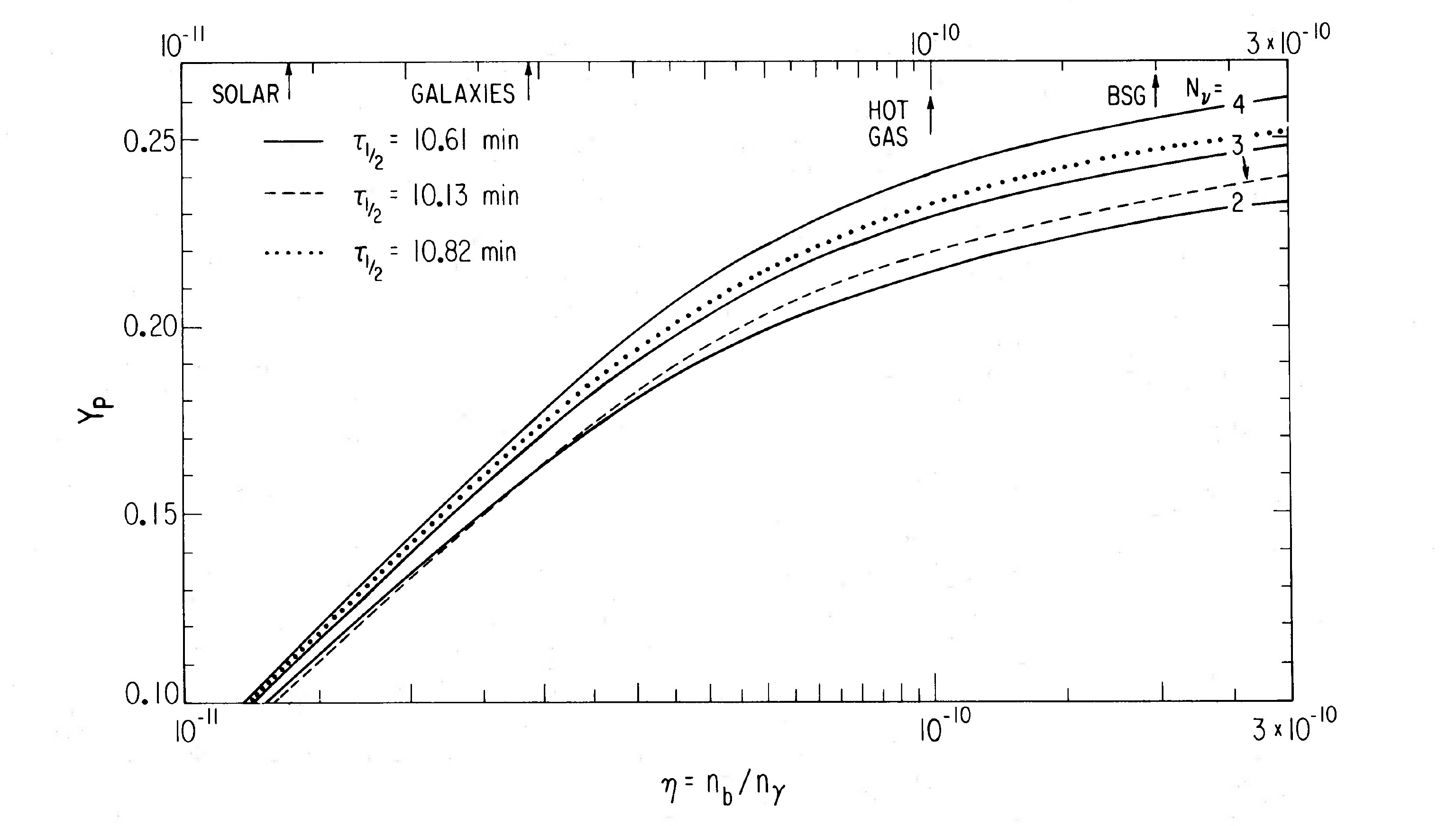}
\caption{
The helium mass fraction as a function of the baryon-to-photon ratio for three choices of $N_\nu = 2, 3$, and 4 and three choices of $\tau_{1/2}$ with $N_\nu = 3$. Figure circa 1981 from ref.~\cite{ossty}. 
\label{fig:ossty}
}
\end{figure}

Historically, progress in BBN has been tightly linked to progress in the measurement of the neutron lifetime.
Until the 1980's, the uncertainties in the prediction of the light element abundances by BBN was  dominated by the uncertainty in these three parameters $(\eta, N_\nu, \tau_n)$; this was prior to the CMB determination of the baryon density, the measurement of the number of neutrino flavors at LEP/SLAC \cite{Janot:2019oyi}, and the vast improvement in the measurement of the neutron lifetime\cite{ossty,ytsso}. For example,  the primordial helium mass fraction is sensitive to the variation in each of these quantities as can be seen in Fig.~\ref{fig:ossty} taken from the 1981 analysis in Ref.~\cite{ossty}. In this figure, the helium mass fraction, $Y_p = \rho(\he4)/\rho_{\rm baryon}$ is plotted as a function of the baryon-to-photon ratio $\eta $ for three values of $N_\nu$ and three values for the neutron lifetime. 
At the time, BBN provided the best estimate for $\eta$ and $N_\nu$ (this was pre-LEP/SLAC and direct limits from accelerators on $N_\nu$ were quite poor \cite{Ellis:1982ej}). The values of $\tau_n$ used here corresponded to what was the ``accepted" value for the neutron half-life, $\tau_{1/2} = 10.61 \pm 0.16$ min  corresponding to a mean life $\tau_n = 918 \pm 14$ s that came from a 1972 paper by Christensen et al. \cite{Christensen:1972pu};  and more recent though very different determinations of $\tau_{1/2} = 10.13 \pm 0.09$ min or $\tau_n = 877 \pm 8$ s by Bondarenko et al. \cite{Bondarenko:1978dn}; and $\tau_{1/2} = 10.82 \pm 0.20$ min or $\tau_n = 937 \pm 17$ s \cite{Byrne:1980vq} by Byrne et al.

It is also interesting to note the range in $\eta$ shown in this 1981 figure. It spans values of $\eta$ from $10^{-11}$ (corresponding to mass-to-light ratios typical of the solar neighborhood) to $3 \times 10^{-10}$. Today, $\eta$ is quite well determined \cite{CFOY,foyy,ysof}: $\eta_{10} = 10^{10} \eta = 6.104 \pm 0.055$ from CMB data alone \cite{Planck2018}, and $\eta_{10} = 6.115 \pm 0.038$ from a combination of BBN, CMB and light element abundance observations.  
Clearly our understanding of the baryon density of the Universe has progressed.

The Review of Particle Properties in 1982 \cite{ParticleDataGroup:1982ifn} quoted an average neutron mean life of $\tau_n = 925 \pm 11$ s
based on the Christensen et al.~and Byrne et al. measurements along with a measurement by Kosvintsev et al. \cite{Kosvintsev:1980uj} which gave $\tau_n = 875 \pm 95$ s. The Bondarenko et al. measurement was excluded as it was ``in significant disagreement with the other two precise direct mean life measurements and the inferred value given by Wilkinson 80'' \cite{Wilkinson:1980ef}. In the 1984 edition of the Review of Particle Properties \cite{ParticleDataGroup:1984mfx}
the Bondarenko measurement was included replacing Kosvintsev 1980
giving $\tau_n = 898 \pm 16$ s, noting that the origin of the discrepancies between the measurements was not known.
Because of the discrepancies, the uncertainty in the mean was inflated by a scale factor of 2.4. 

It wasn't until 1990, when Mampe et al. \cite{Mampe:1989xx} presented a significantly more accurate result $\tau_n = 887.6 \pm 3.0$ s using ultra cold neutrons in a fluid-walled bottle. This had a strong effect on the world average which was now $888.6 \pm 3.5$ based on seven measurements. The discrepancy was largely gone and the scale factor for the uncertainty was reduced to 1.3,
in part due to the withdrawal of the 1980 measurement by Byrne et al. 

\begin{figure}[!htb]
\vskip -1in
\includegraphics[width=0.80\textwidth]{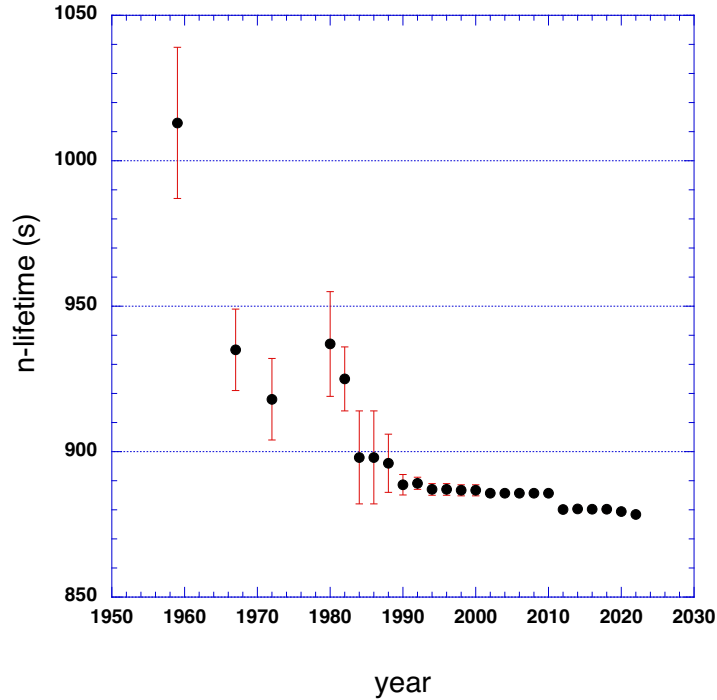}
\vskip -1in
\caption{
The average mean lifetime of the neutron as compiled by the Review of Particle Properties/Physics.
Note that the uncertainty in the mean life for more recent measurements is smaller than the symbol showing the mean. 
\label{fig:ave}
}
\end{figure}

The evolution of the world average is plotted in Fig.~\ref{fig:ave} showing the average mean lifetime since 1960.
One clearly sees the marked drop in the value and uncertainty following the Mampe et al. measurement. The impact of this measurement on BBN was immediate \cite{ossw} since, as described in more detail below, the uncertainty in the predicted helium abundance is very sensitive to the uncertainty in the neutron mean lifetime \cite{yof},
\beq
\label{eq:dYp}
\frac{\Delta Y_p}{Y_p} \simeq 0.730 \frac{\Delta \tau_n}{\tau_n} \, ,
\eeq
meaning that a drop in the uncertainty from 16 to 3.5 reduced the uncertainty in $Y_P$ from 0.0032 to 0.0007 assuming $Y_p = 0.247$. 

By the time of the 2002 edition of the Review of Particle Physics \cite{ParticleDataGroup:2002ivw}
there were enough high quality direct measurements allowing the Particle Data Group (PDG) to drop measurements with uncertainties larger than 10s. This led to another impressive drop in the uncertainty of the mean, with $\tau_n = 885.7 \pm 0.8$ s. This drop was largely driven by the measurement of Arzumanov et al. \cite{Arzumanov:2000ma} and further dropped the uncertainty in $Y_P$ to 0.0002 making the dominant source of uncertainty (at the time) other nuclear rates \cite{cfo1}. 

At about this time, WMAP provided the first all-sky measurements of anisotropies in the CMB temperature  \cite{WMAP:2003elm}.  The these fluctuations encode a wealth of cosmological parameters, including the baryon density or baryon-to-photon ratio $\eta$.  Using this input from the CMB  along with the improvements in the measurements of $\tau_n$ effectively made BBN a parameter-free theory \cite{cfo2}. 

\begin{figure}[!htb]
\includegraphics[width=0.8\textwidth]{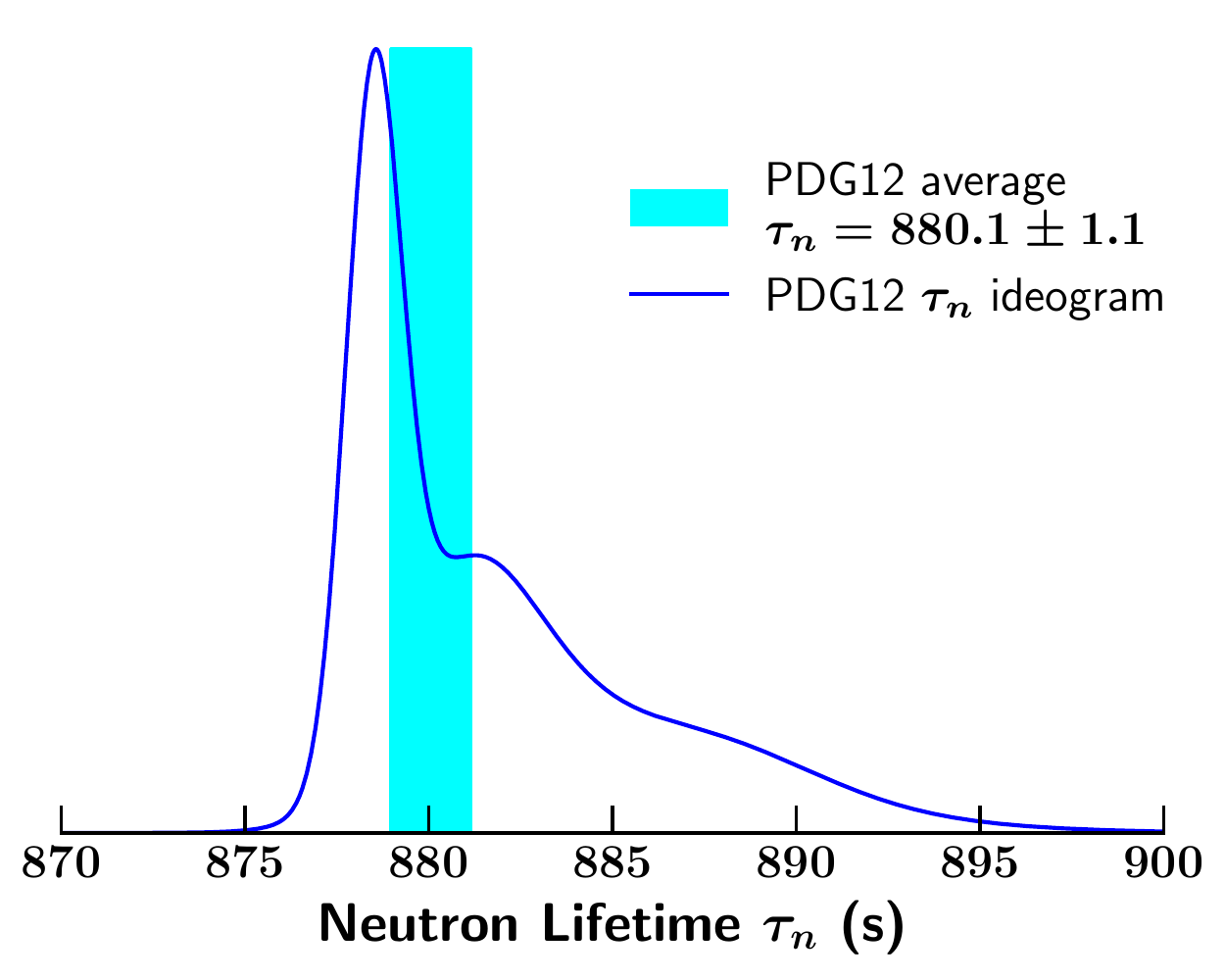}
\caption{
Ideogram for the seven measurements contributing to the PDG average neutron mean life in 2012. 
\label{ideo1}
}
\end{figure}

However it is certainly reasonable to question whether the quoted uncertainties are truly reflective of systematic errors included. In 2005, Serebrov et al. \cite{Serebrov:2004zf} published a result which was in fact systematically lower the previous world average, $\tau_n = 878.5 \pm 0.7 \pm 0.3$. This value was used in a BBN analysis in Ref.~\cite{grant}.  As a sole and severely discrepant value, it was not included in the PDG average until 2012 \cite{ParticleDataGroup:2012pjm}
when additional low values were reported. The 2012 average was now $\tau_n = 880.1 \pm 1.1$ where the uncertainty has been scaled by a factor of 1.8. This seismic shift in $\tau_n$ is clearly seen in Fig.~\ref{fig:ave}. An ideogram of the seven measurements leading to this average is shown in Fig.~\ref{ideo1}.
Here, we see the onset of a discrepancy in $\tau_n$. The vertical blue rectangle corresponds to the 2012 mean and 1$\sigma$ spread. 
A review including a discussion of the methods used to obtain $\tau_n$ can be found in \cite{Wietfeldt:2011suo}. 

\begin{figure}[!htb]
\includegraphics[width=0.8\textwidth]{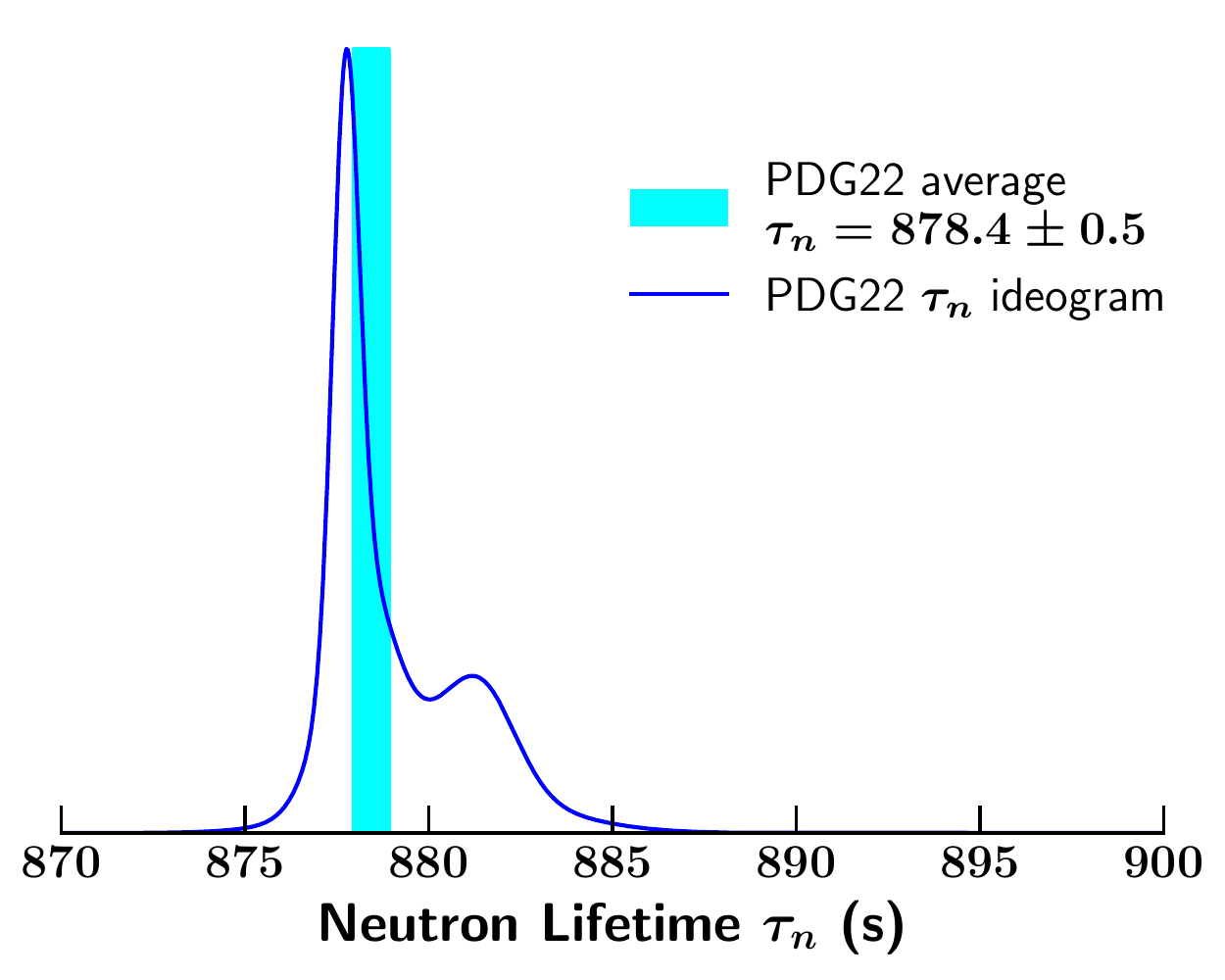}
\caption{
Ideogram for the eight measurements contributing to the PDG average neutron mean life in 2022. 
\label{ideo2}
}
\end{figure}

Subsequent changes in the world average have been relatively minor--certainly from the perspective of BBN calculations~\cite{ysof}, 
where the current value  of 
\beq
\tau_n = 878.4 \pm 0.5 {\rm s}
\label{tauncurr}
\eeq
based on eight measurements \cite{Serebrov:2004zf,Pichlmaier:2010zz,Steyerl:2012zz,Arzumanov:2015tea,Serebrov:2017bzo,Pattie:2017vsj,Ezhov:2014tna,UCNt:2021pcg,ParticleDataGroup:2022pth} is used. As the uncertainties in individual measurements continue to drop, the discrepancy appears more pronounced and the uncertainty already includes a scale factor of 1.8 \cite{ParticleDataGroup:2022pth}.
This can be seen in the current ideogram shown in Fig.~\ref{ideo2}.

Despite the tightness of the world average in~Eq.~(\ref{tauncurr}), the neutron lifetime remains an outstanding puzzle.  
The results used in Eq.~(\ref{tauncurr}) are based on confined ultracold neutrons, with earlier experiments using material bottles, and later experiments using magnetic and gravitational traps.  The measurements in traps give consistent results, with recent very precise UCN$\tau$ determination \cite{UCNt:2021pcg} dominating the weighted average among them and leading to the sharp peak in Fig.~\ref{ideo2}. The most recent bottle measurements give lifetimes systematically longer than the trap measurements, leading to the shoulder to the right of the peak in Fig.~\ref{ideo2}.  The resulting {\em dispersion} between the trap and bottle measurements is of order $\Delta \tau_n \sim 5 \ \rm sec$, and could stem from systematic differences between the methods.  
There is in addition an in-beam measurement by Yue et al. \cite{Yue:2013qrc} with a relatively high value of $\tau_n = 887.7 \pm 1.2 \pm 1.9$ s. This differs from the best trap measurement by $\Delta \tau_n \sim  10 \ \rm sec$.
We will not here enter further into the current discrepancy, noting only that for a recent discussion of the differences between the ultracold neutron results (the ones used in the PDG average) and the in-beam results, see \cite{Czarnecki:2018okw} as well as other articles in this Special Issue.

In what follows, we will first briefly review the role of the neutron lifetime in standard BBN calculations in section \ref{SBBN}. We will also review the current results of BBN based on the latest input nuclear cross-sections (including $\tau_n$) and abundance data.
In section \ref{sensitivity}, we will discuss the sensitivity of the results to $\tau_n$. We will also test the potential effect of the $\tau_n$ discrepancy by instead of Monte Carlo sampling of the Gaussian corresponding to the mean value of $\tau_n$ (and its uncertainty), but as a test, we will use the ideogram in Fig.~\ref{ideo2} to sample values of $\tau_n$ - keeping in mind that this procedure is not fully rigorous. In section \ref{predictions}, we will take a novel approach to $\tau_n$ and treat it again as a parameter for 
which we generate a likelihood function and make a BBN prediction for $\tau_n$.
Finally in section \ref{summary}
we summarize the current state of BBN with respect to the neutron lifetime.

\section{Standard BBN}
\label{SBBN}

Standard BBN (SBBN) is built upon 
 Standard Model of nuclear and particle physics, in the background of the Friedmann-Robertson-Walker cosmological model based on Einstein gravity. 
 We assume only 
the standard set of nuclear and particle interactions and nuclear and particle content\footnote{In fact, we assume $\Lambda$CDM, so that in addition to these Standard Model particles and interactions, there is (1) a nonzero cosmological constant $\Lambda$ which will be totally negligible during BBN, and (2) cold dark matter which we take to be so weakly interacting as to have no effect on BBN.  These assumptions can be relaxed; see reviews in refs.~\cite{Pospelov:2010hj,Jedamzik:2009uy,Malaney:1993ah}.}, in particular, with $N_\nu = 3$.
Furthermore, in SBBN, we assume a radiation dominated Universe  
during the epoch of nucleosynthesis. The radiation density can be expressed as
\begin{equation}
\rho = {\pi^2 \over 30} \left( 2 + {7 \over 2} + {7 \over 4}N_\nu \right) T^4 ,
\label{rho}
\end{equation}
taking into account the contributions of photons, electrons and positrons, and 
neutrino flavors appropriate for temperatures $T > 1$ MeV.
The expansion rate of the Universe, is determined by the Hubble parameter which can be expressed as
\beq
H^2 = \frac{8\pi}{3} G_{\rm N} \rho \, ,
\label{hubb}
\eeq
where $G_{\rm N}$ is Newton's constant and scales as $H \propto G_{\rm N}^{1/2} T^2$
in a radiation dominated universe. 

\begin{figure}[!htb]
\includegraphics[width=1\textwidth]{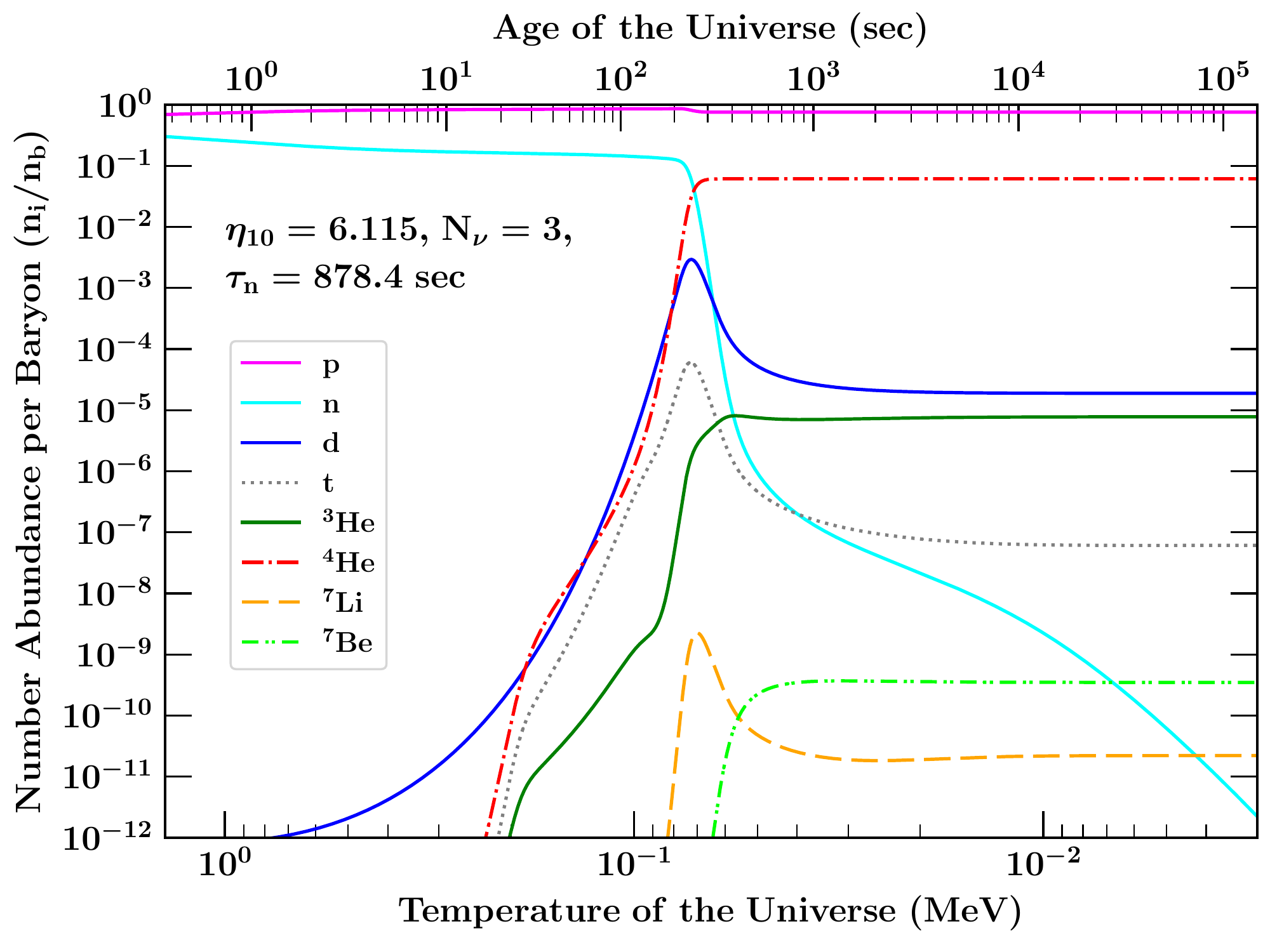}
\caption{
Time evolution of the light element abundances during BBN. Note that time (upper axis) increases to the right, and so the temperature is shown decreasing to the right.
\label{fig:timeevol}
}
\end{figure}

The cosmic evolution of the light nuclides is plotted in Fig.~\ref{fig:timeevol}; many critical features
can be understood analytically, as we now summarize.
At temperatures $T \gtrsim 1$ MeV,  weak
interactions between neutrons and protons maintain equilibrium. These are:
\begin{eqnarray}
\label{eq:n-p}
n + e^+ & \leftrightarrow & p + \bar{\nu}_e  \nonumber \\
p + e^- & \leftrightarrow & n + \nu_e  \nonumber \\
n & \leftrightarrow & p + e^- \bar{\nu}_e \, .
\end{eqnarray}
As one might expect, the weak interaction rates scale as $\Gamma_{\rm wk} \propto G_{\rm F}^2 T^5$, where $G_{\rm F}$ is the Fermi constant. 
These reactions freeze-out when their interaction rates become slower than the expansion rate of the Universe determined by the Hubble parameter, or in other words, when the mean time between interactions is longer than the age of the Universe, determined by $H^{-1}$. 
Thus, the freeze-out condition is set by
\beq
G_{\rm F}^2 T^5 \sim \Gamma_{\rm wk} (T_{\rm f})  = H(T_{\rm f}) \sim G_{\rm N}^{1/2} T^2 \, .
\label{fo}
\eeq

Weak freeze-out occurs at $T_{\rm f} \simeq 0.84$~MeV, and corresponds to a flattening of the neutron and proton curves in Fig.~\ref{fig:timeevol}.
At freeze-out, the neutron-to-proton ratio is given
approximately by the Boltzmann factor,
$(n/p)_{\rm f} \simeq e^{-\Delta m/T_{\rm f}} \sim 1/4.66$, where $\Delta m=m_n-m_p = 1.293$ MeV is the neutron--proton
mass difference. The resulting mass fraction of neutrons at freezeout is thus $X_{n,\rm f} = \left. n/(n+p) \right|_{\rm f} = (n/p)_{\rm f}/[1+(n/p)_{\rm f}]$.  

After freeze-out, free neutron decays reduce the ratio
slightly, which leads to the shallow slope in the neutron and proton curves in Fig.~\ref{fig:timeevol}.  Accounting for these decays, 
$X_{n,\rm BBN} = X_{n,\rm f} e^{-t_{\rm BBN}/\tau_n} \simeq [n/(n+p)]_{\rm f} e^{-t_{\rm BBN}/\tau_n}$ 
before nucleosynthesis begins at time $t_{\rm BBN}$. 
For $t_{\rm BBN} = 322$ s determined by the time when the photo-destruction rates of deuterium freeze-out, and $\tau_n = 878.4$ s, we have $X_{n,\rm BBN} = 1/8.17$  or $(n/p)_{\rm BBN} = 1/7.17$.
A useful and more elaborate semi-analytic description
of freeze-out can be found in~\cite{Bernstein1988,muk}.

At $t_{\rm BBN}$, nucleosynthesis activity reaches a crescendo.  Deuterons are formed via $n p \rightarrow d \gamma$, and then undergo a series of  strong reactions culminating in \he4 production.  These rapid reactions lead to a dramatic drop in the free $n$ abundance, as seen in Fig.~\ref{fig:timeevol}.  Finally, the strong reactions also freeze out, largely due to the inability to overcome the Coulomb barrier in the expanding and cooling plasma.   Then the stable nuclides plateau, while the radioactive species ultimately decay--as most evidently seen by the free neutron decay in Fig.~\ref{fig:timeevol} at late times.

\begin{figure}[!htb]
\includegraphics[width=0.80\textwidth]{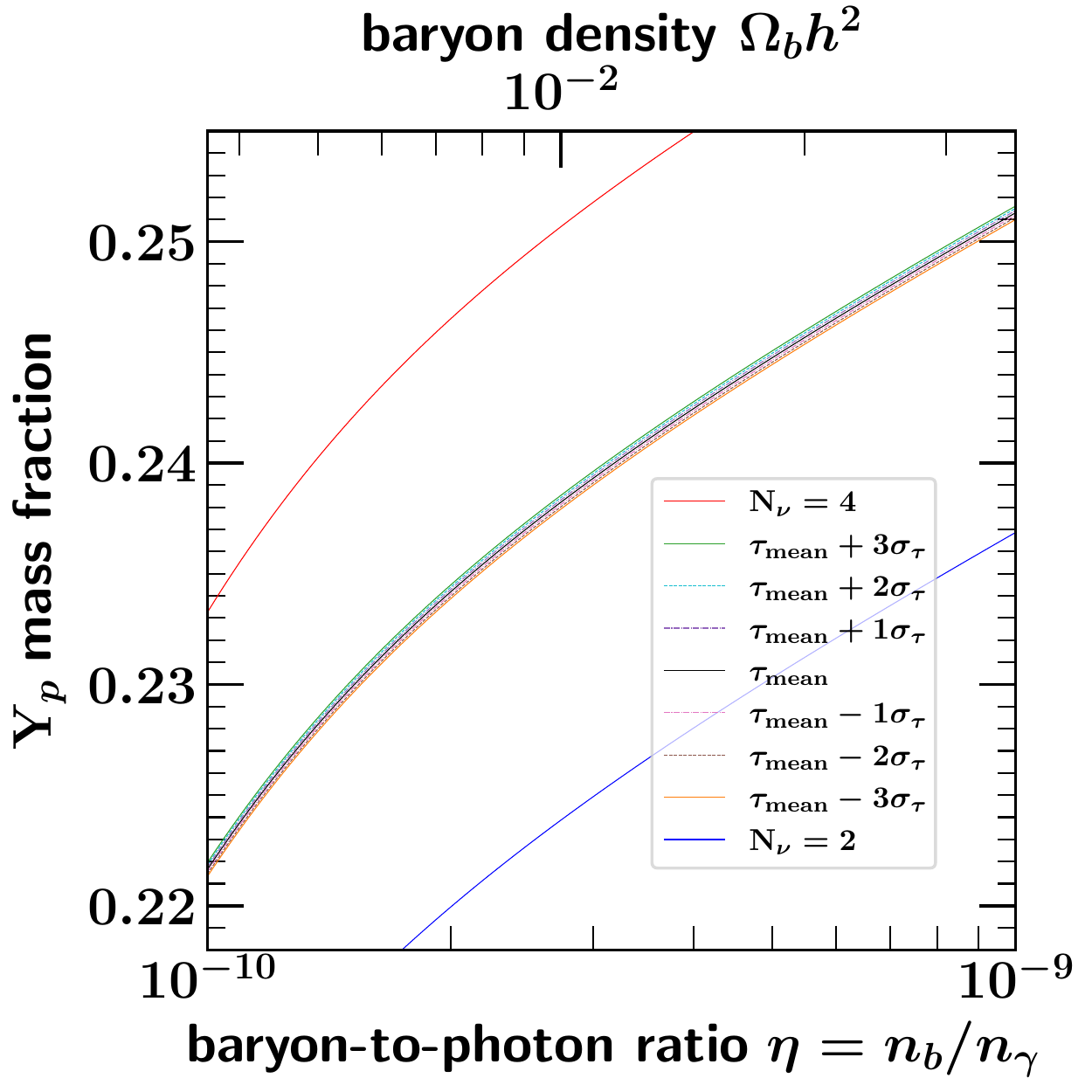}
\caption{
The helium mass fraction as a function of the baryon-to-photon ratio for three choices of $N_\nu = 2, 3$, and 4 and for a spread in values of $\tau_{1/2}$ up to $\pm 3 \sigma$ about the mean with $N_\nu = 3$. 
\label{fig:ytsso}
}
\end{figure}

The neutron lifetime plays two roles in BBN calculations.
First, it is used to normalize the zero-temperature matrix element for weak $n-p$ interconversions (\ref{eq:n-p}).
The rates for these reactions
scale as $\Gamma_{n \leftrightarrow p} \propto 1/\tau_n$. Thus $\tau_n$ affects the determination of the freeze-out temperature of these weak interactions. Freeze-out is determined from the competition between the weak interaction rates and the Hubble expansion as in Eq.~\ref{fo}. Thus an increase in the neutron lifetime leads to an increase in the freeze-out temperature
leading to more neutrons at freeze-out (recall the n/p ratio scales as $e^{-\Delta m/T}$ prior to freeze-out) and hence more \he4 as seen in Figs.~\ref{fig:ossty} and \ref{fig:ytsso}.

Secondly the neutron lifetime controls the rate of free neutron decay, which 
occurs between weak freezeout at $t_{\rm f} \sim 1 \ \rm sec$
and the end of the D bottleneck at $t_d \approx 320 \ \rm sec$.
These decays lower the $n/p$ ratio. So an increase in the neutron lifetime leads again to more neutrons left over and hence a higher \he4 abundance.

Similar arguments can be made for the dependence of $Y_p$ on $N_\nu$ and $\eta$. As one can see from Fig.~\ref{fig:ytsso}, $Y_p$ increases with increases in all three inputs: $\eta$, $N_\nu$, and $\tau_n$.  It is interesting to compare Fig.~\ref{fig:ytsso}
with Fig.~\ref{fig:ossty}.
Qualitatively, they are similar.
However there are several jarring differences. First, as remarked earlier, the domain in $\eta$, previously did not even cover the current best fit of $\eta_{10} = 6.115$. Second, as has been stressed in the Introduction, the relevant values of $\tau_n$ are very different. 
Whereas in Fig.~\ref{fig:ossty}
values of $\tau_n$ between 877 s and 937 s, were deemed reasonable. 
In Fig.~\ref{fig:ytsso},
the 3$\sigma$ spread in curves shown for $N_\nu = 3$, cover 876.8 s to 879.8 s. Finally, whilst the uncertainty in 
$Y_p$ from $N_\nu$ (for $\Delta N_\nu = \pm 1$)  and $\tau_n$ were comparable in Fig.~\ref{fig:ossty}, as one can plainly see, for the same spread in $N_\nu$, the uncertainty due a $3 \sigma$ spread in $\tau_n$ is almost negligible.

\section{Abundance sensitivities to $\tau_n$}
\label{sensitivity}

In this section, we review the dependence of the light element abundances to $\tau_n$. As we have discussed earlier, we expect 
the abundance of \he4 to increase with increasing $\tau_n$. 
It is instructive and relatively straightforward to estimate this dependence. As is well known, the dominant isotope emerging from BBN is \he4 with a mass fraction of roughly 25\%. In contrast, the abundance (by number) of D and \he3 is only of order $10^{-5}$. Therefore to a good approximation, 
we can assume that after the deuterium bottleneck, all neutrons will eventually go to \he4 and we can write
\beq
Y_p = 2X_{n,\rm BBN} = 2 \left. \frac{n/p}{1+n/p} \right|_{\rm BBN}
\eeq
 for the \he4 mass fraction, where the second and third expressions are evaluated at 
 the time $t_{\rm BBN}$ at the end of the deuterium bottleneck.
Recalling that the neutron-to-proton ratio at freeze-out is fixed by the Boltzmann factor $e^{-\Delta m/T_{\rm f}}$, when BBN begins, the $n/p$ ratio is lowered by neutron decays so that 
\begin{eqnarray}
    Y_p & \approx & 2  (X)_{T_{\rm f}} \ e^{-t_{\rm BBN}/\tau_{\rm n}}  \approx 2 \frac{e^{-\Delta m/T_{\rm f}}}{1+e^{-\Delta m/T_{\rm f}}} \ e^{-t_{\rm BBN}/\tau_{\rm n}}    
\end{eqnarray}
which explicitly shows the direct $\tau_n$ dependence as well as its indirect influence
through the freezeout temperature $T_{\rm f}$.  The latter is set by the 
equating the weak interaction rates 
$\Gamma_{np} \propto \tau_n^{-1} T^5$ 
and the Hubble
rate $H \sim G_N^{1/2} T^2$, which gives
\begin{equation}
    T_{\rm f} \propto \tau_n^{1/3}
\end{equation}
We then can trace the effect of $\tau_n$ shifts by looking at the linearized response to changes in the neutron lifetime:
\begin{eqnarray}
    \frac{\Delta Y_p}{Y_p} & \approx & \left( \frac{\partial Y_p}{\partial X} \frac{\partial X}{\partial(n/p)} \frac{\partial(n/p)}{\partial T_{\rm f}}  \frac{\partial T_{\rm f}}{\partial \tau_n} +\frac{\partial Y_p}{\partial \tau_n} \right)  \frac{\Delta \tau_n}{Y_p} \, .
\end{eqnarray}
Then using $\partial Y_p/\partial X = 2$, $\partial X/\partial (n/p) = 1/(1+n/p)^2$, $\partial (n/p)/\partial T_{\rm f} = (n/p) \Delta m/T_{\rm f}^2$, we can write
\begin{eqnarray}
  \frac{\Delta Y_p}{Y_p}   & \approx & \left( \frac{1}{3} \frac{\Delta m}{T_{\rm f }(1 + n/p)_{\rm f}} +\frac{t_{\rm BBN}}{\tau_n} \right) \frac{\Delta \tau_n}{\tau_n} \,.
\end{eqnarray}
Finally using $T_{\rm f} = 0.84$ MeV, $(n/p)_{\rm f} = 1/4.66$, $t_{\rm BBN} = 322$, and $\tau_n = 878.4$ s, we have
\beq
 \frac{\Delta Y_p}{Y_p} \approx (0.42 + 0.37)  \frac{\Delta \tau_n}{\tau_n} = 0.79  \frac{\Delta \tau_n}{\tau_n} \, ,
\eeq
 which can be compared with the full numerical dependence (at the best fit value of $\eta$) of \cite{foyy}
given in Eq.~(\ref{eq:dYp}).
Recall that the contributions are comparable and work in the same direction:  a longer $\tau_n$ leads to more \he4 
due to (1)
a earlier weak freezeout and (2) more free neutrons surviving decay.

The numerical scaling of the light element abundances to $\eta$, $N_\nu$ and $\tau_n$, is
\beqar
Y_p &=& 0.2467\!\left(\frac{\e10}{6.115}\right)^{0.040}\!\!\left(\frac{N_\nu}{3.0}\right)^{0.163}\!\!\left(\frac{\tau_n}{878.4 s}\right)^{0.730} \, ,
\label{yfit}
\eeqar
\beqar
\frac{\rm D}{\rm H} &=& 2.496\!\times\! 10^{-5}\!\left(\frac{\e10}{6.115}\right)^{-1.634}\!\!\left(\frac{N_\nu}{3.0}\right)^{0.405}\!\!\left(\frac{\tau_n}{878.4 s}\right)^{0.413} \, ,
\eeqar
\beqar
\frac{\he3}{\rm H} &=& 1.041\!\times\! 10^{-5}\!\left(\frac{\e10}{6.115}\right)^{-0.570}\!\!\left(\frac{N_\nu}{3.0}\right)^{0.138}\!\!\left(\frac{\tau_n}{878.4 s}\right)^{0.127} \, ,
\eeqar
\beqar
\frac{\li7}{\rm H} &=& 4.937\!\times\! 10^{-10}\!\left(\frac{\e10}{6.115}\right)^{2.117}\!\!\left(\frac{N_\nu}{3.0}\right)^{-0.285}\!\!\left(\frac{\tau_n}{878.4 s}\right)^{0.431} \, , 
\label{li7fit}
\eeqar
where other inputs such as nuclear cross section have been set at their mean values. These fits are normalized at the Standard Model value of $N_\nu = 3$, the best fit value of $\eta_{10} = 6.115$ \cite{ysof} and Eq.~(\ref{tauncurr}) for $\tau_n$.

\begin{figure}[!htb]
\includegraphics[width=0.90\textwidth]{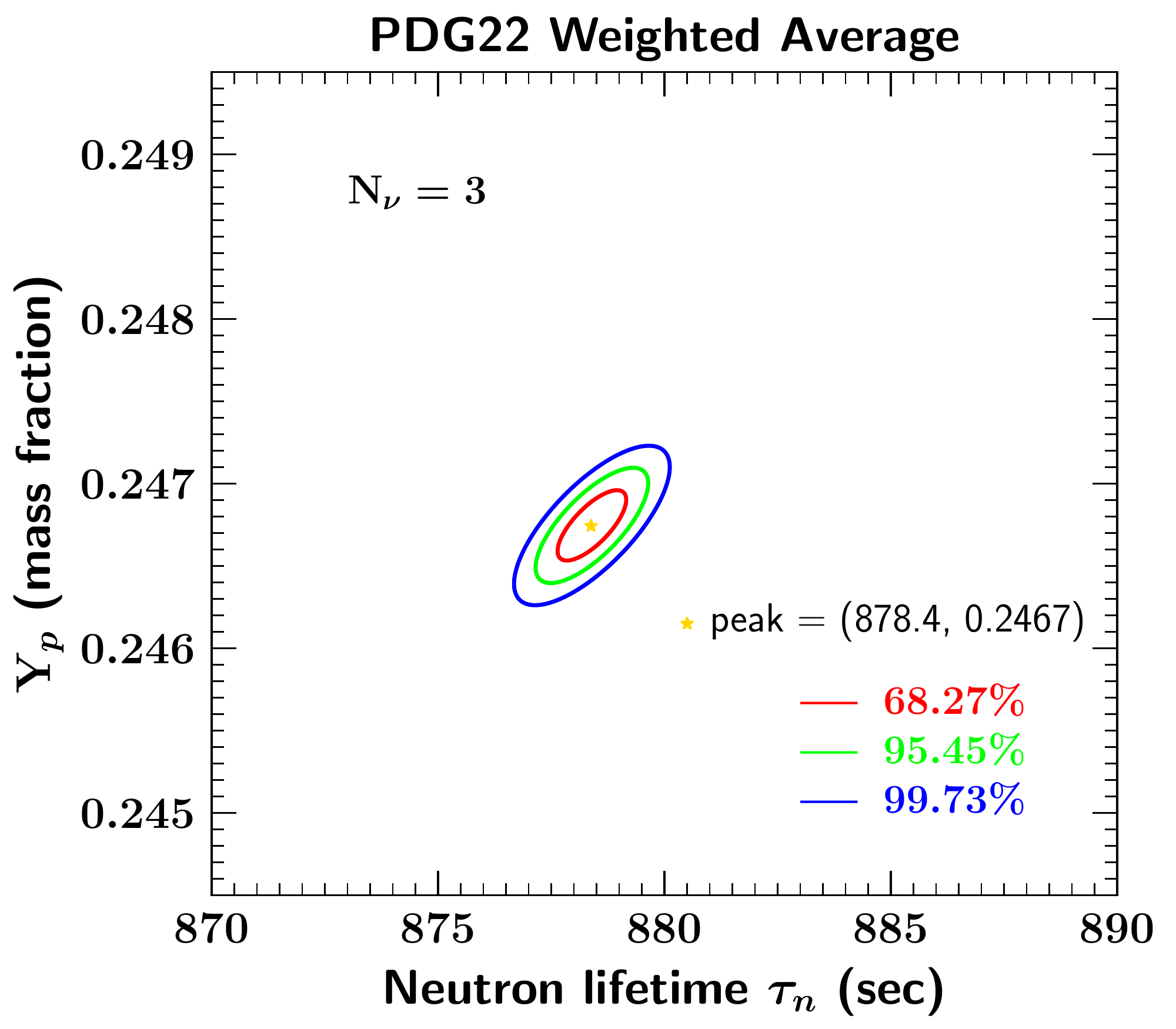}
\caption{The sensitivity of the \he4 abundance to the neutron
mean lifetime, assuming a Gaussian distribution for $\tau_n$ with mean and uncertainty given by Eq.~(\ref{tauncurr}). In addition to the peak of the likelihood, denoted by a star, we show the 1, 2 and 3$\sigma$ contours. }
\label{ytau}
\end{figure}

The sensitivity of the \he4 abundance to the neutron mean life is shown in Fig.~\ref{ytau}.
Typically, because the uncertainty in the neutron mean life is small, BBN predictions for the light element abundances are obtained by combining the BBN likelihood function, $ {\mathcal L}_{\rm BBN}(\eta;X_i)$, 
where the abundances $X_i$ cover D, \he3, \he4 and \li7, with a CMB likelihood function taken from Planck data \cite{Planck2018}, $  {\mathcal L}_{\rm CMB}(\eta,Y_p)$, where we include the dependence on $Y_p$ and do not assume any a priori relation between $Y_p$ and $\eta$ \cite{CFOY,foyy}. Throughout, we are assuming $N_\nu = 3$. We can, however, include the explicit dependence on $\tau_n$ to form the following likelihood function
\beq
{\mathcal L}_{{\rm BBN+CMB}+\tau_n}(\tau_n,X_i) \propto \int 
  {\mathcal L}_{\rm CMB}(\eta,Y_p) \
  {\mathcal L}_{\rm BBN}(\eta, \tau_n;X_i){\mathcal L}_{\tau_n}( \tau_n) \ d\eta \, ,
\label{CMB+BBN+taun}
\eeq
where ${\mathcal L}_{\tau_n}( \tau_n)$ can either be derived from the Gaussian with
mean and uncertainty given in Eq.~(\ref{tauncurr}) or the ideogram in Fig.~\ref{ideo2}.
The projection of this likelihood function (for $X_i = Y_p$) onto the $(\tau_n, Y_p)$ plane is shown in Fig.~\ref{ytau}.
Here, we  show the 1, 2, and 3$\sigma$ contours from a Monte Carlo scan over the neutron mean life assuming Eq.~(\ref{tauncurr})
with a Gaussian distributed uncertainty. The star indicates the peak value of the likelihood function. The tightness of the ellipses is a consequence of the small uncertainty in $\tau_n$. This plot is an update of that in \cite{CFOY}. 
The expected correlation between the neutron mean lifetime and \he4 abundance prediction is clear. It is not perfectly linear because other reaction rate uncertainties significantly contribute to the total uncertainty in \he4 as sampled in our Monte Carlo. Note that marginalizing over $\tau_n$ determines the  theoretical \he4 likelihood function discussed below.

\begin{figure}[!htb]
\includegraphics[width=0.90\textwidth]{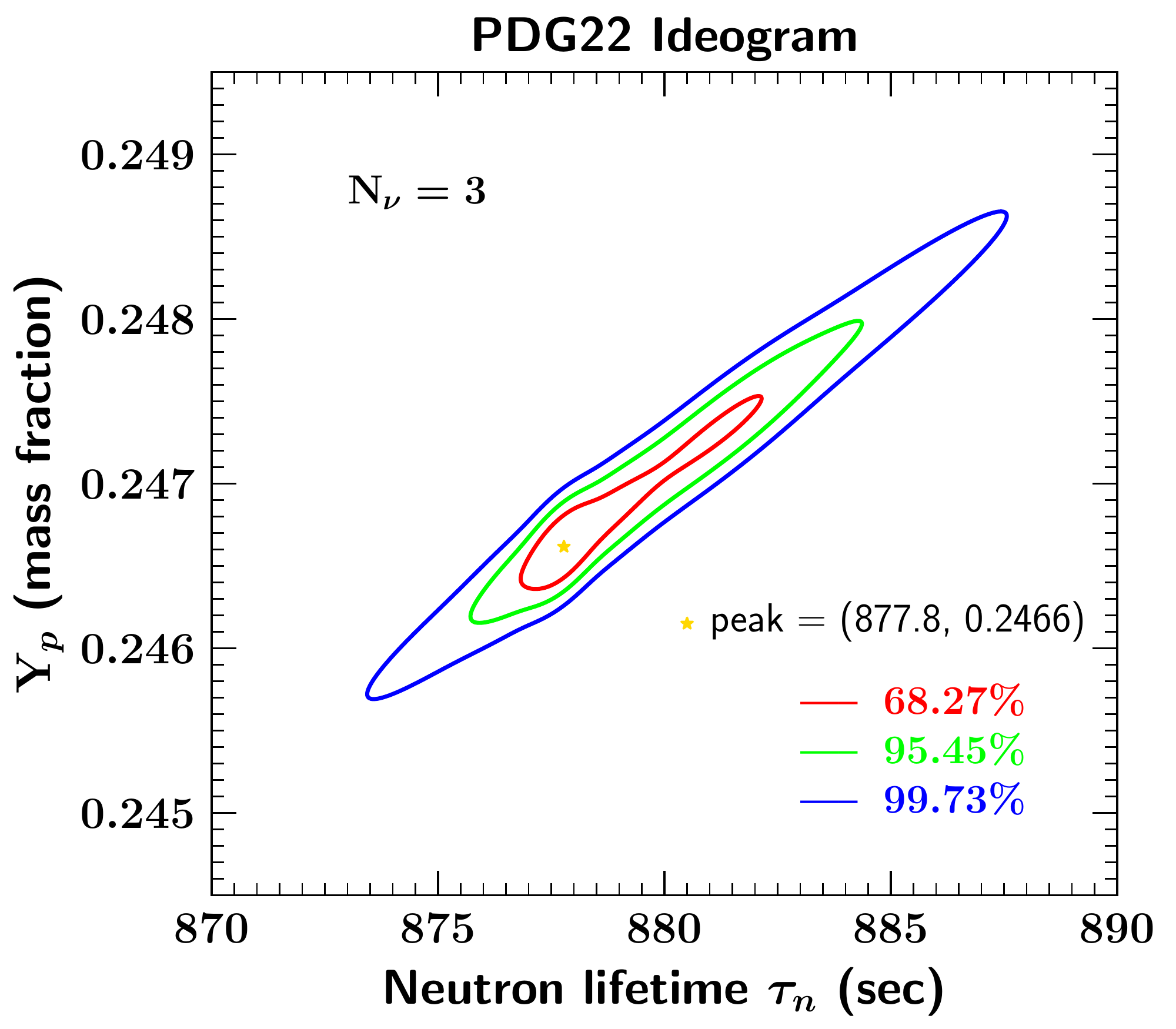}
\caption{As in Fig.~\ref{ytau} where we have assumed the distribution of $\tau_n$ taken from the ideogram in Fig.~\ref{ideo2}. }
\label{ytaui}
\end{figure}

If instead of taking a Gaussian 
distribution with mean and uncertainty given by Eq.~(\ref{tauncurr}), we use the distribution indicated by the ideogram in Fig.~\ref{ideo2}, we obtain the elongated prediction for $Y_p$ shown in Fig.~\ref{ytaui}. While the ideogram distribution should not be taken as statistically rigorous, we remind the reader that there is somewhat significant dispersion in the experimental results for $\tau_n$ (a scale factor of 1.8 was already applied to obtain $\sigma_{\tau_n} = 0.5$ s)
and the Gaussian may mask the true uncertainty. \footnote{Including the in-beam measurement would further increase the dispersion requiring a scale factor of 2.2.} The tail end of the ideogram tends up to lifetimes of about 885 s, leading to significantly more \he4.

\begin{figure}[!htb]
\includegraphics[width=0.90\textwidth]{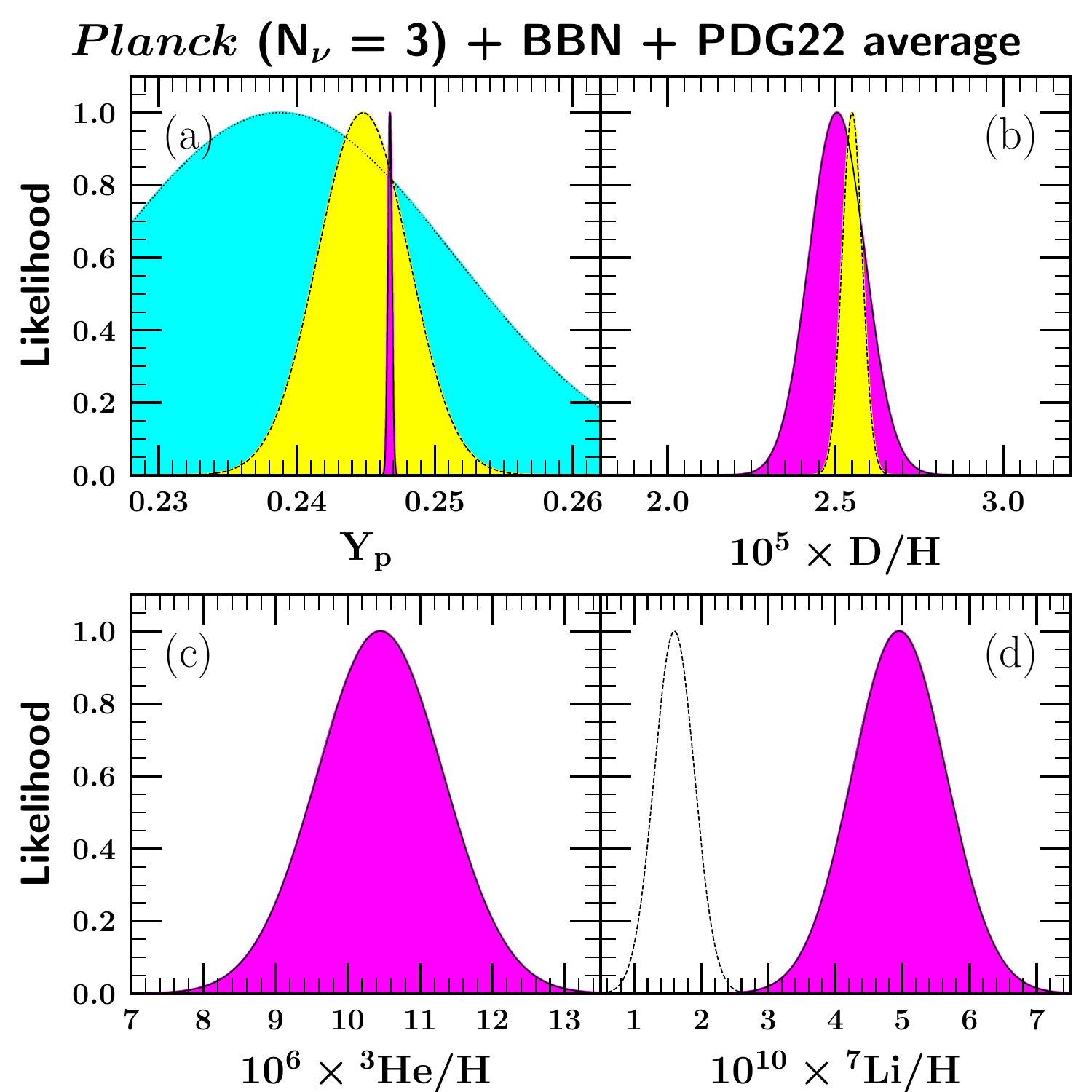}
\caption{Light element abundance likelihood functions.  Shown are likelihoods for each of the light nuclides.  The solid-lined,
dark-shaded (purple) curves are 
the BBN+CMB predictions, based on {\em Planck} inputs as discussed
in the text.  The dashed-lined, light-shaded 
(yellow) curves show astronomical measurements
of the primordial abundances, for all but \he3 where 
reliable primordial abundance measures do not exist.  For 
\he4, the dotted-lined, medium-shaded (cyan) curve shows
the independent CMB determination of \he4. 
\label{fig:2x2abs_2d}}
\end{figure}

It is also common to compare the predictions of BBN to the observations by examining the theoretical and observational likelihood functions \cite{CFOY,foyy,yof}. The theoretical likelihood function can be expressed as a convolution of the BBN theory, dependent on $\eta$ (we again fix $N_\nu = 3$)
and the CMB likelihood functions. The combined likelihood is defined by
\beq
{\mathcal L}_{\rm BBN+CMB}(X_i) \propto \int 
  {\mathcal L}_{\rm CMB}(\eta,Y_p) \
  {\mathcal L}_{\rm BBN}(\eta;X_i) \ d\eta \propto \int {\mathcal L}_{{\rm BBN+CMB}+\tau_n}(\tau_n,X_i) d\tau_n \, .
\label{CMB+BBN}
\eeq
The second proportionality indicates that the combined BBN and CMB likelihood function is given by marginalizing Eq.~(\ref{CMB+BBN+taun}) over $\tau_n$. 
We normalize each of the likelihood functions so that at their peaks $\mathcal{L}=1$.
Figure~\ref{fig:2x2abs_2d} shows
the comparison of these likelihood functions for (a) $Y_p$ (upper left), (b) D/H (upper right), (c) \he3/H (lower left), and (d) \li7/H (lower right). In the case of \he4, we show all three likelihood functions. The combined CMB-BBN likelihood from Eq. (\ref{CMB+BBN}), ${\mathcal L}_{\rm CMB-BBN}(Y)$, is shaded purple. The observational likelihood determined from \cite{Aver:2020fon,Aver:2021rwi}
\beq
\label{eq:Ypobs}
Y_{p,\rm obs} = 0.2448 \pm 0.0033 \, .
\eeq
is shaded yellow. The CMB-only likelihood, given by
\beq
{\mathcal L}_{\rm CMB}(Y_p) \propto \int 
  {\mathcal L}_{\rm CMB}(\eta,Y_p) \ d\eta \, .
  \label{CMByp}
\eeq
 is shaded cyan. Given the observational and CMB uncertainties, the agreement is quite good.

In the case of D/H, the observational likelihood is determined from \cite{pc,cooke,riemer,bala,cookeN,riemer17,zava,CPS}
\beq \label{dhobs}
\left(\frac{\rm D}{\rm H}\right)_{\rm obs} = (2.55 \pm 0.03) \times 10^{-5} \, .
\eeq
The agreement between the CMB and BBN likelihoods as seen in panel b) of Fig.~\ref{fig:2x2abs_2d}
is a major success for early Universe cosmology.  We have not shown the observational likelihood function for \he3 as there are no unambiguous observations which can be associated with the primordial abundance. Similarly, it has been argued that the long-standing problem associated with \li7 \cite{cfo5}, is no longer well-founded \cite{Fields:2022mpw}. The main arguments for associating the observational abundance with the primordial abundance relied on the unambiguous observation of \li6 in halo stars, as well as the lack of dispersion in \li7 abundances at low metallicity. Both of these arguments are now suspect. We show nevertheless,
the \li7 abundance from low metallicity by the unshaded likelihood curve at \li7/H $= (1.6 \pm 0.3) \times 10^{-10}$ \cite{ryan2000}. 

The CMB-BBN likelihoods in Fig.~\ref{fig:2x2abs_2d}
are summarized by
the predicted abundances 
\beqar
Y_p &=& 0.2467\pm0.0002 \qquad \qquad \qquad (0.2467) \label{meyp} \\
{\rm D/H} &=& (2.506\pm0.083)\times 10^{-5} \qquad (2.505 \times 10^{-5})\label{medh}\\
\he{3}/{\rm H} &=& (10.45\pm 0.87)\times 10^{-6} \qquad (10.45 \times 10^{-6}) \label{mehh} \\
\li{7}/{\rm H} &=& (4.96\pm 0.70)\times 10^{-10} \qquad (4.95 \times 10^{-10}) \label{melh}
\eeqar
where the central values give the mean,
and the error gives the $1\sigma$ variance.
The final number in parentheses gives the value at the peak of the distribution.

\begin{figure}[!htb]
\includegraphics[width=0.90\textwidth]{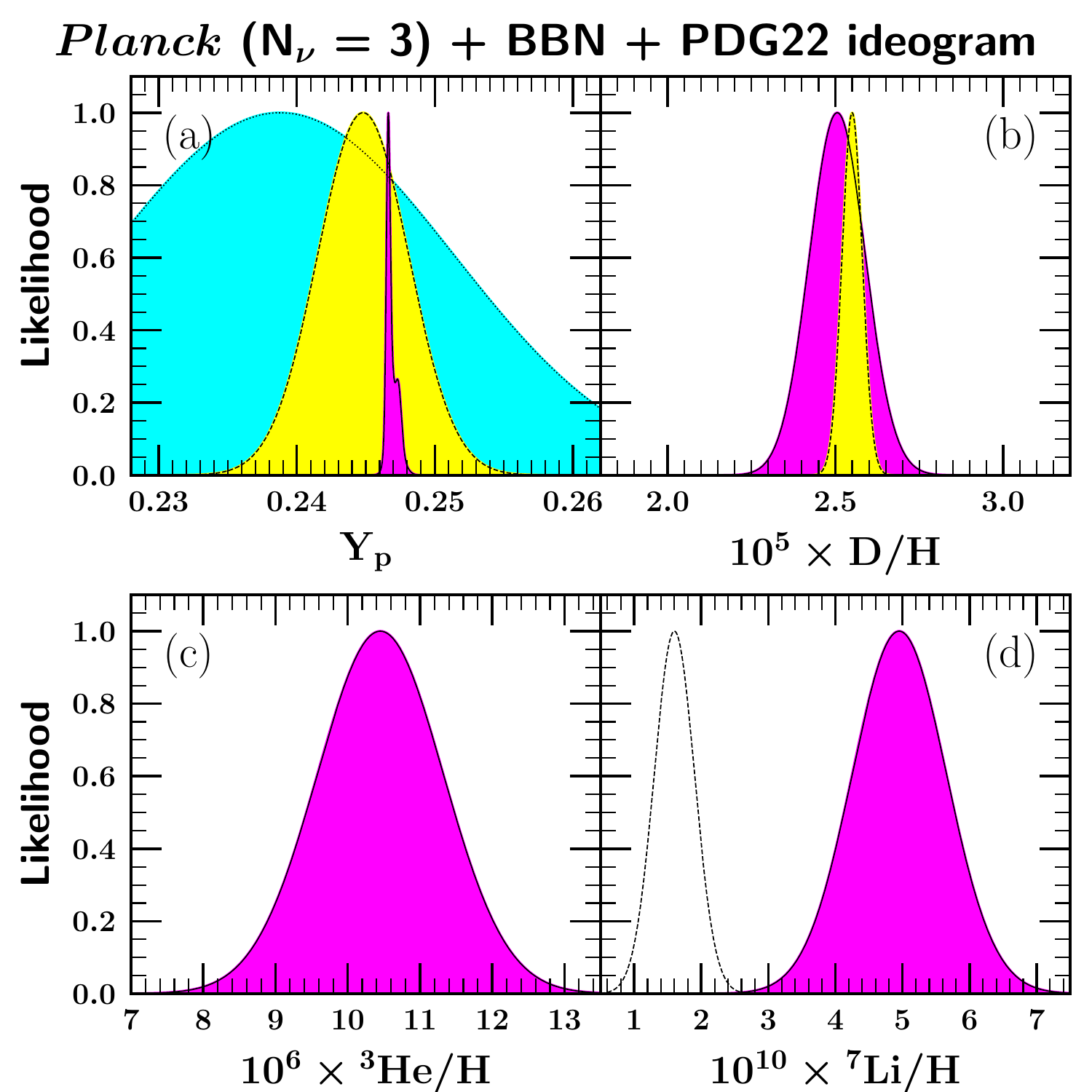}
\caption{As in Fig.~\ref{fig:2x2abs_2d}, where the neutron mean life is selected from the ideogram in Fig.~\ref{ideo2} rather than the Gaussian distribution. 
\label{fig:2x2abs_2di}}
\end{figure}

For comparison, in Fig.~\ref{fig:2x2abs_2di}, we also show the same likelihood functions for each of the light elements, but instead in our Monte Carlo choose values of the neutron mean life from the ideogram in Fig.~\ref{ideo2}, rather than the Gaussian distribution. As one can see, that apart from the feature on the high side of the BBN \he4 distribution (purple shaded likelihood in panel a), the results are very similar which emphasizes the lack of sensitivity to the current neutron mean life given the experimental uncertainties, even with the dispersion among recent measurements.

Similarly, the CMB-BBN likelihoods in Fig.~\ref{fig:2x2abs_2di}
using the ideogram in Fig.~\ref{ideo2} can be  summarized by
the predicted abundances 
\beqar
Y_p &=& 0.2469\pm0.0004 \qquad \qquad \qquad (0.2466) \label{meyp2} \\
{\rm D/H} &=& (2.507\pm0.083)\times 10^{-5} \qquad (2.505 \times 10^{-5})\label{medh2}\\
\he{3}/{\rm H} &=& (10.45\pm 0.87)\times 10^{-6} \qquad (10.45 \times 10^{-6}) \label{mehh2} \\
\li{7}/{\rm H} &=& (4.96\pm 0.70)\times 10^{-10} \qquad (4.95 \times 10^{-10}) \,.
\label{melh2}
\eeqar
The most striking effect between the two distributions, is the increase by a factor of 2
in the predicted uncertainty in $Y_p$.

Finally, we consider the case in which the $\tau_n$ errors are dominated by systematics.  In this case, it would be inappropriate to use the world average procedure we and the PDG have adopted.  Using Eq.~(\ref{eq:dYp}) with $\Delta \tau_n = 5$ s gives $\Delta Y_p = 0.0010 \sim 0.3 \sigma_{\rm obs}(Y_p)$, where $\sigma_{\rm obs}(Y_p)$ is the observed \he4 error in eq.~(\ref{eq:Ypobs}).  We see that while this shift is large compared to the $Y_p$ theoretical error in eq.~(\ref{meyp}) and (\ref{meyp2}), it is small compared to the observed uncertainty.  
Even if we consider a $\Delta t_n = 10 \ \rm s$ discrepancy between the  in-beam measurement and UCN$\tau$ trap measurements, the \he4 change is $\Delta Y_p = 0.0021 = 0.6 \sigma_{\rm obs}(Y_p)$, and again the shift in the prediction is modest compare to the observed \he4 errors.  The lesson is that the neutron lifetime does not impact the basic BBN concordance, but also that future improvements in both $Y_p$ and $\tau_n$ will make this comparison more interesting, as we now see.

\section{BBN `predictions' of $\tau_n$}
\label{predictions}

\begin{figure}[!htb]
\includegraphics[width=0.90\textwidth]{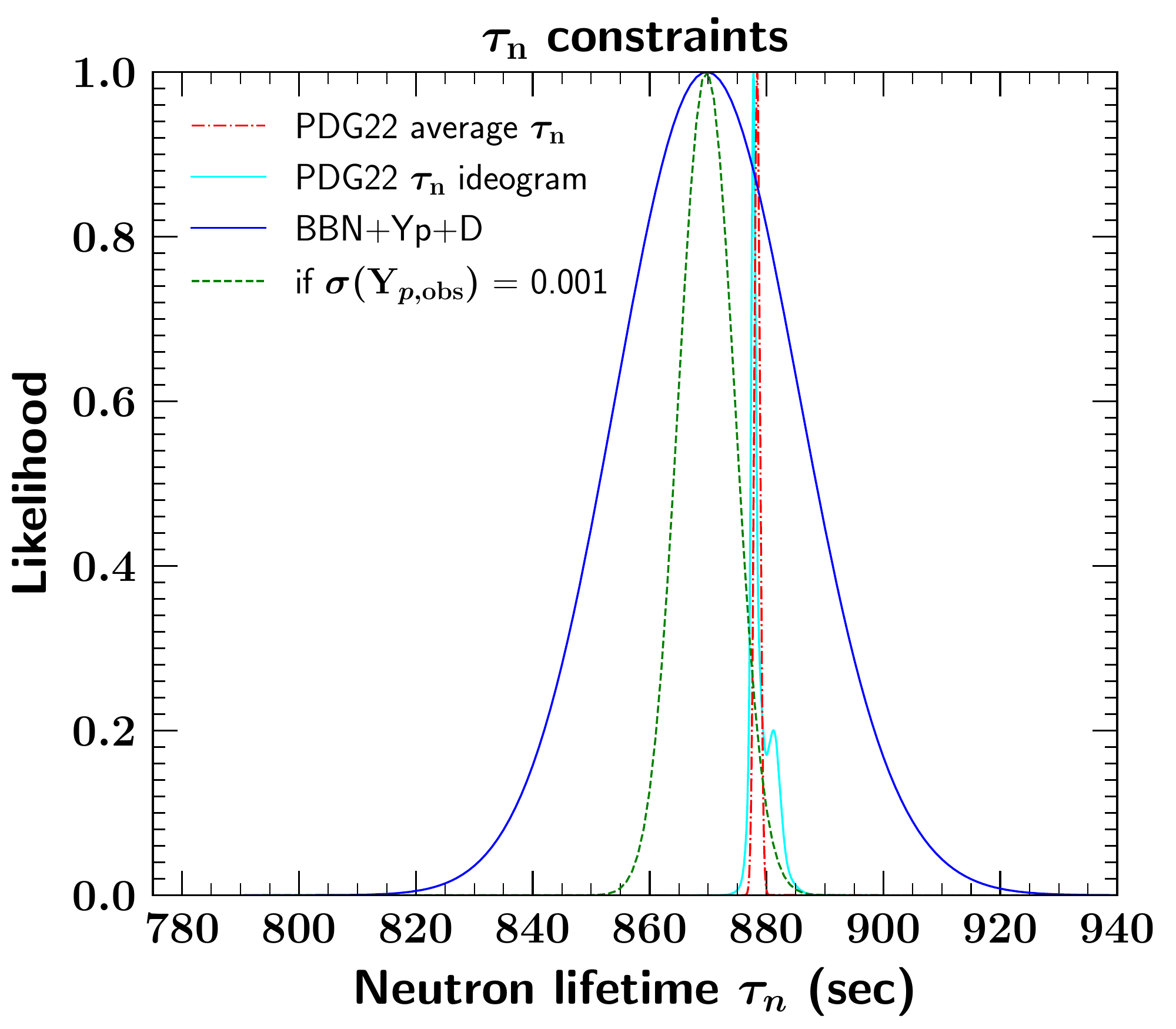}
\caption{Comparison of the current and potential future BBN predicted likelihood function for $\tau_n$ with the experimental likelihood. The solid blue curve shows the current BBN prediction which is tightened to the green dashed curve if the uncertainty in the helium mass fraction can be reduced to $\sigma_{Y_p} = 0.001$. For comparison, the red dot dashed curve shows the experimental value for $\tau_n$ represented by a Gaussian. The distribution represented by the ideogram is shown by the thin cyan solid curve. 
}
\label{Ltaun}
\end{figure}

It is also possible to
combine the likelihood function in Eq.~(\ref{CMB+BBN+taun}) with an observational likelihood function $\mathcal{L}_{\rm OBS}(Y_p)$ from Eq.~(\ref{eq:Ypobs}) to obtain a likelihood function for $\tau_n$
\beq
\mathcal{L}_{\tau_n}(\tau_n) \propto \int  {\mathcal L}_{\rm CMB}(\eta,Y_p) \
  {\mathcal L}_{\rm BBN}(\eta, \tau_n;Y_p) \mathcal{L}_{\rm OBS}(Y_p) d\eta dY_p \, .
\eeq
A similar exercise was performed in \cite{Salvati:2015wxa}. 
This likelihood function (again normalized so that the peak is at $\mathcal{L} = 1$) is shown in Fig.~\ref{Ltaun}
by the blue curve. 
It is characterized by an approximate Gaussian given by 
\beq
\label{eq:tau-predicted}
\tau_n({\rm BBN+CMB}) = 870.2 \pm 15.8 {\rm s} \qquad \qquad \mbox{for}~ \  \sigma_{Y_p} = 0.0033 \, .
\eeq
In Fig.~\ref{Ltaun}, we compare this BBN `prediction' with the Gaussian given by Eq.~\ref{tauncurr} (red dot-dashed curve), and that given by the ideogram (thin cyan curve).

Although we did not discuss it here, the combined BBN and CMB analysis is capable of setting strong constraints on the number of relativistic degrees of freedom at the time of BBN characterized by $N_\nu$. 
Currently the joint analysis gives \cite{ysof} $N_\nu = 2.989 \pm 0.141$ or a $2 \sigma$ upper limit $N_\nu < 3.180$. The S4 CMB-only sensitivity \cite{CMB-S4:2016ple} is expected to reach $\sigma(N_\nu) = 0.07$. That means that if BBN is to remain competitive in the determination of $N_\nu$, improvements in \he4 observations are needed so that the uncertainty in $Y_p$ is lowered to 0.001. In that case, we obtain a better determination of $\tau_n$ shown by the green dashed curve in Fig.~\ref{Ltaun}. 
In this case, we have
\beq
\tau_n({\rm BBN+CMB}) = 869.8 \pm 4.8 {\rm s} \qquad \qquad \mbox{for}~ \ \sigma_{Y_p} = 0.001 \, .
\eeq
This can be compared to likelihood from Eq.~(\ref{tauncurr}) shown by the dot-dashed red curve.  It is no surprise that it will be very difficult for BBN + CMB to compete with direct experimental measurements of the neutron mean life.

\section{Summary and Outlook}
\label{summary}

Motivated by recent interest in the neutron lifetime, we have studied the effect of $\tau_n$ and its uncertainties on BBN in the post-{\em Planck} era of precision cosmology. In BBN calculations, the neutron lifetime normalizes the $n \leftrightarrow p$ interconversions prior to weak freezeout, and then controls free neutron decay.  The current discrepancy in neutron lifetime variations thus has an impact on cosmology, which we have assessed.

The continued improvements in $\tau_n$ measurements have played a key factor in the present robustness and precision of BBN.  In the 1980s the neutron lifetime was a limiting factor in BBN predictions, and it remains the case that the $\tau_n$ uncertainty dominates the uncertainty in the \he4 predictions.  But as we have shown, modern $\tau_n$ data lead to tight predictions of the light element abundances--even given the systematic uncertainties in the neutron lifetime measurements.  We have shown that combining neutron lifetime measurements as in the PDG ideogram somewhat broadens the errors in the \he4 prediction, but this change remains far below the observed \he4 errors.  Even a $\Delta \tau_n = 10 \ \rm sec$ systematic uncertainty would change the $Y_p$ prediction by less than the $1\sigma$ uncertainty in the astronomical observations, and the change in the other light elements is negligible. This robustness of the BBN predictions represents a triumph of nuclear experiment, and allows BBN to play a key role in cosmology and particle physics.

Indeed, the precision of BBN nuclear inputs--and that of CMB observations--allows us to turn the problem around and make a joint BBN+CMB prediction of the neutron lifetime.  Our result, $\tau_n({\rm BBN+CMB}) = 870.2 \pm 15.8 \ \rm sec$ (eq.~\ref{eq:tau-predicted}) is consistent within errors with all experimental determinations.  This concordance represents a success of the big bang cosmology, but the agreement arises because the errors in the BBN+CMB predictions are too large (at present) to discriminate among the experiments.  We can compare this cosmological measurement of the neutron lifetime with another recent astrophysical determination based on the NASA {\em Lunar Prospector} measurements of the neutron flux around the Moon:  $\tau_{n}^{\rm Moon} = 887 \pm 14_{\rm stat}{}^{+7}_{-3 \; \rm syst} \ \rm sec$ \cite{Wilson:2020lfr}.  We see that the two independent results are consistent within errors, and that interestingly the errorbars are almost identical.  We note that unlike the lunar result, the cosmologically-preferred central value lies somewhat below the laboratory results, though the offset is not statistically significant.

Upcoming experiments and observations can improve the current situation and sharpen the link between the neutron lifetime and cosmology.  This is important not only for finding the resolution of the $\tau_n$ puzzle, but also for BBN.  As cosmological measurements become even more precise, it remains important to improve the determination of $\tau_n$ and other nuclear inputs to the BBN predictions.  
We look forward to progress on several fronts.
\begin{itemize}
    \item New neutron lifetime measurements are planned.  These include (a) both an upgrade magneto-gravitational trap experiment UCN$\tau$+, and (b) an upgraded pulsed beam experiment, Beam Lifetime 3 (BL3) \cite{Wietfeldt:2022tma}.  These can shed new light on and perhaps resolve the $\tau_n$ puzzle.

\item
    The next generation CMB measurements from CMB-S4 will significantly improve both the determination of $Y_p$ and $N_\nu$ from the CMB \cite{CMB-S4:2016ple}.   Improved $\tau_n$ measurements will be important for BBN to fully exploit these results, particularly $N_\nu$.

\item
    The ongoing effort to improve astronomical $Y_p$ determinations continues.  As we have discussed here and elsewhere \cite{ysof}, reaching the ambitious goal of $\sigma_{\rm obs}(Y_p) = 0.001$ would open a new window on new physics generally and $\tau_n$ in particular.

\end{itemize}
As these new measurements come in, they will help deepen the links among nuclear physics, particle physics, and early Universe cosmology.

\vspace{2em}

{\em Acknowledgments:}  We are grateful for illuminating discussions with Chen-Yu Liu regarding neutron lifetime measurements. TRIUMF receives federal funding via a contribution agreement with the National Research Council of Canada. The work of K.A.O. is supported in part by DOE grant DE-SC0011842 at the University of
Minnesota.


\begin{thebibliography}{0}

\bibitem{bbn} T.~P.~Walker, G.~Steigman, D.~N.~Schramm, K.~A.~Olive and H.~S.~Kang,
  Astrophys.\ J.\  {\bf 376} (1991) 51;
K.~A.~Olive, G.~Steigman and T.~P.~Walker,
  Phys.\ Rept.\  {\bf 333}, 389 (2000)
  [astro-ph/9905320];
B.~D.~Fields and K.~A.~Olive,
Nucl. Phys. {\bf A777}, 208 (2006);
B.~D.~Fields, P.~Molaro and S.~Sarkar,
  Chin.\ Phys.\ C {\bf 38} (2014)
  [arXiv:1412.1408 [astro-ph.CO]];
  G.~Steigman,
  Ann.\ Rev.\ Nucl.\ Part.\ Sci.\  {\bf 57}, 463 (2007)
  [arXiv:0712.1100 [astro-ph]].

\bibitem{iocco}
  F.~Iocco, G.~Mangano, G.~Miele, O.~Pisanti and P.~D.~Serpico,
  Phys.\ Rept.\  {\bf 472}, 1 (2009)
  [arXiv:0809.0631 [astro-ph]].

\bibitem{coc18}
   C.~Pitrou, A.~Coc, J.~P.~Uzan and E.~Vangioni,
  Phys.\ Rept.\  {\bf 754}, 1 (2018)
  [arXiv:1801.08023 [astro-ph.CO]].

  \bibitem{CFOY}
R.~H.~Cyburt, B.~D.~Fields, K.~A.~Olive and T.-H. Yeh,
  Rev.\ Mod.\ Phys.\  {\bf 88}, 015004 (2016)
  [arXiv:1505.01076 [astro-ph.CO]].

  \bibitem{foyy}
B.~D.~Fields, K.~A.~Olive, T.~H.~Yeh and C.~Young,
JCAP \textbf{03}, 010 (2020)
[arXiv:1912.01132 [astro-ph.CO]].



\bibitem{ParticleDataGroup:2022pth}
R.~L.~Workman \textit{et al.} [Particle Data Group],
PTEP \textbf{2022}, 083C01 (2022).

\bibitem{Planck2018}
N.~Aghanim \textit{et al.} [Planck Collaboration],
Astron. Astrophys. \textbf{641} (2020), A6
[arXiv:1807.06209 [astro-ph.CO]].

\bibitem{Janot:2019oyi}
P.~Janot and S.~Jadach,
Phys. Lett. B \textbf{803}, 135319 (2020)
[arXiv:1912.02067 [hep-ph]].

\bibitem{ossty}
K.~A.~Olive, D.~N.~Schramm, G.~Steigman, M.~S.~Turner and J.~M.~Yang,
Astrophys. J. \textbf{246}, 557 (1981)

\bibitem{ytsso}
J.~M.~Yang, M.~S.~Turner, G.~Steigman, D.~N.~Schramm and K.~A.~Olive,
Astrophys. J. \textbf{281}, 493-511 (1984)

\bibitem{Ellis:1982ej}
J.~R.~Ellis and K.~A.~Olive,
Nucl. Phys. B \textbf{223}, 252-268 (1983).

\bibitem{Christensen:1972pu}
C.~J.~Christensen, A.~Nielsen, A.~Bahnsen, W.~K.~Brown and B.~M.~Rustad,
Phys. Rev. D \textbf{5}, 1628-1640 (1972)

\bibitem{Bondarenko:1978dn}
L.~N.~Bondarenko, V.~V.~Kurguzov, Y.~A.~Prokofev, E.~V.~Rogov and P.~E.~Spivak,
Pisma Zh. Eksp. Teor. Fiz. \textbf{28}, 328-333 (1978)

\bibitem{Byrne:1980vq}
J.~Byrne, J.~Morse, K.~F.~Smith, F.~Shaikh, K.~Green and G.~L.~Greene,
Phys. Lett. B \textbf{92}, 274-278 (1980)

\bibitem{ysof}
T.~H.~Yeh, J.~Shelton, K.~A.~Olive and B.~D.~Fields,
JCAP \textbf{10}, 046 (2022)
[arXiv:2207.13133 [astro-ph.CO]].

\bibitem{ParticleDataGroup:1982ifn}
M.~Roos \textit{et al.} [Particle Data Group],
Phys. Lett. B \textbf{111}, 1-294 (1982)

\bibitem{Kosvintsev:1980uj}
Y.~Y.~Kosvintsev, Y.~A.~Kushnir, V.~I.~Morozov and G.~I.~Terekhov,
JETP Lett. \textbf{31}, 236 (1980)


\bibitem{Wilkinson:1980ef}
D.~H.~Wilkinson,
Prog. Part. Nucl. Phys. \textbf{6}, 325-332 (1981)
doi:10.1016/0146-6410(81)90041-7

\bibitem{ParticleDataGroup:1984mfx}
C.~G.~Wohl \textit{et al.} [Particle Data Group],
Rev. Mod. Phys. \textbf{56}, S1-S304 (1984)

\bibitem{Mampe:1989xx}
W.~Mampe, P.~Ageron, C.~Bates, J.~M.~Pendlebury and A.~Steyerl,
Phys. Rev. Lett. \textbf{63}, 593-596 (1989)

\bibitem{ossw}
K.~A.~Olive, D.~N.~Schramm, G.~Steigman and T.~P.~Walker,
Phys. Lett. B \textbf{236}, 454-460 (1990)

\bibitem{yof}
T.~H.~Yeh, K.~A.~Olive and B.~D.~Fields,
JCAP \textbf{03}, 046 (2021)
doi:10.1088/1475-7516/2021/03/046
[arXiv:2011.13874 [astro-ph.CO]].

\bibitem{ParticleDataGroup:2002ivw}
K.~Hagiwara \textit{et al.} [Particle Data Group],
Phys. Rev. D \textbf{66}, 010001 (2002)

\bibitem{Arzumanov:2000ma}
S.~Arzumanov, L.~Bondarenko, S.~Chernavsky, A.~Fomin, V.~Morozov, Y.~Panin, W.~Drexel, K.~Schreckenbach, P.~Geltenbort and J.~Pendlebury,
Phys. Lett. B \textbf{483}, 15-22 (2000)

  \bibitem{cfo1}
R.~H.~Cyburt, B.~D.~Fields and K.~A.~Olive,
{\it New Astron.\  } {\bf 6} (1996) 215
[arXiv:astro-ph/0102179].

\bibitem{WMAP:2003elm}
D.~N.~Spergel \textit{et al.} [WMAP],
Astrophys. J. Suppl. \textbf{148}, 175-194 (2003)
[arXiv:astro-ph/0302209 [astro-ph]].


 \bibitem{cfo2}
R.~H.~Cyburt, B.~D.~Fields and K.~A.~Olive,
Astropart.\ Phys.\  {\bf 17} (2002) 87
[arXiv:astro-ph/0105397].

\bibitem{Serebrov:2004zf}
A.~Serebrov, V.~Varlamov, A.~Kharitonov, A.~Fomin, Y.~Pokotilovski, P.~Geltenbort, J.~Butterworth, I.~Krasnoschekova, M.~Lasakov and R.~Tal'daev, \textit{et al.}
Phys. Lett. B \textbf{605}, 72-78 (2005)
[arXiv:nucl-ex/0408009 [nucl-ex]].

\bibitem{grant}
G.~J.~Mathews, T.~Kajino and T.~Shima,
Phys. Rev. D \textbf{71}, 021302 (2005)
[arXiv:astro-ph/0408523 [astro-ph]].

\bibitem{ParticleDataGroup:2012pjm}
J.~Beringer \textit{et al.} [Particle Data Group],
Phys. Rev. D \textbf{86}, 010001 (2012)

\bibitem{Wietfeldt:2011suo}
F.~E.~Wietfeldt and G.~L.~Greene,
Rev. Mod. Phys. \textbf{83}, no.4, 1173-1192 (2011)

\bibitem{Pichlmaier:2010zz}
A.~Pichlmaier, V.~Varlamov, K.~Schreckenbach and P.~Geltenbort,
Phys. Lett. B \textbf{693}, 221-226 (2010)

\bibitem{Steyerl:2012zz}
A.~Steyerl, J.~M.~Pendlebury, C.~Kaufman, S.~S.~Malik and A.~M.~Desai,
Phys. Rev. C \textbf{85}, 065503 (2012)

\bibitem{Arzumanov:2015tea}
S.~Arzumanov, L.~Bondarenko, S.~Chernyavsky, P.~Geltenbort, V.~Morozov, V.~V.~Nesvizhevsky, Y.~Panin and A.~Strepetov,
Phys. Lett. B \textbf{745}, 79-89 (2015)

\bibitem{Serebrov:2017bzo}
A.~P.~Serebrov, E.~A.~Kolomensky, A.~K.~Fomin, I.~A.~Krasnoshchekova, A.~V.~Vassiljev, D.~M.~Prudnikov, I.~V.~Shoka, A.~V.~Chechkin, M.~E.~Chaikovskiy and V.~E.~Varlamov, \textit{et al.}
Phys. Rev. C \textbf{97}, no.5, 055503 (2018)
[arXiv:1712.05663 [nucl-ex]].

\bibitem{Pattie:2017vsj}
R.~W.~Pattie, Jr., N.~B.~Callahan, C.~Cude-Woods, E.~R.~Adamek, L.~J.~Broussard, S.~M.~Clayton, S.~A.~Currie, E.~B.~Dees, X.~Ding and E.~M.~Engel, \textit{et al.}
Science \textbf{360}, no.6389, 627-632 (2018)
[arXiv:1707.01817 [nucl-ex]].

\bibitem{Ezhov:2014tna}
V.~F.~Ezhov, A.~Z.~Andreev, G.~Ban, B.~A.~Bazarov, P.~Geltenbort, A.~G.~Glushkov, V.~A.~Knyazkov, N.~A.~Kovrizhnykh, G.~B.~Krygin and O.~Naviliat-Cuncic, \textit{et al.}
JETP Lett. \textbf{107}, no.11, 671-675 (2018)
[arXiv:1412.7434 [nucl-ex]].

\bibitem{UCNt:2021pcg}
F.~M.~Gonzalez \textit{et al.} [UCN\ensuremath{\tau}],
Phys. Rev. Lett. \textbf{127}, no.16, 162501 (2021)
[arXiv:2106.10375 [nucl-ex]].

\bibitem{Yue:2013qrc}
A.~T.~Yue, M.~S.~Dewey, D.~M.~Gilliam, G.~L.~Greene, A.~B.~Laptev, J.~S.~Nico, W.~M.~Snow and F.~E.~Wietfeldt,
Phys. Rev. Lett. \textbf{111}, no.22, 222501 (2013)
[arXiv:1309.2623 [nucl-ex]].

\bibitem{Czarnecki:2018okw}
A.~Czarnecki, W.~J.~Marciano and A.~Sirlin,
Phys. Rev. Lett. \textbf{120}, no.20, 202002 (2018)
[arXiv:1802.01804 [hep-ph]].


\bibitem{Pospelov:2010hj}
M.~Pospelov and J.~Pradler,
Ann. Rev. Nucl. Part. Sci. \textbf{60}, 539-568 (2010)
[arXiv:1011.1054 [hep-ph]].

\bibitem{Jedamzik:2009uy}
K.~Jedamzik and M.~Pospelov,
New J. Phys. \textbf{11}, 105028 (2009)
[arXiv:0906.2087 [hep-ph]].

\bibitem{Malaney:1993ah}
R.~A.~Malaney and G.~J.~Mathews,
Phys. Rept. \textbf{229}, 145-219 (1993)


\bibitem{Bernstein1988}
J.~Bernstein, L.~S.~Brown and G.~Feinberg,
Rev. Mod. Phys. \textbf{61}, 25 (1989)

\bibitem{muk}
V.~F.~Mukhanov,
  Int.\ J.\ Theor.\ Phys.\  {\bf 43}, 669 (2004)
  [astro-ph/0303073].

\bibitem{Aver:2020fon}
E.~Aver, D.~A.~Berg, K.~A.~Olive, R.~W.~Pogge, J.~J.~Salzer and E.~D.~Skillman,
JCAP \textbf{03}, 027 (2021)
[arXiv:2010.04180 [astro-ph.CO]].

\bibitem{Aver:2021rwi}
E.~Aver, D.~A.~Berg, A.~S.~Hirschauer, K.~A.~Olive, R.~W.~Pogge, N.~S.~J.~Rogers, J.~J.~Salzer and E.~D.~Skillman,
Mon. Not. Roy. Astron. Soc. \textbf{510}, no.1, 373-382 (2021)
[arXiv:2109.00178 [astro-ph.GA]].


 \bibitem{pc}
  M.~Pettini and R.~Cooke,
  Mon.\ Not.\ Roy.\ Astron.\ Soc.\  {\bf 425}, 2477 (2012)
  [arXiv:1205.3785 [astro-ph.CO]].

    \bibitem{cooke}
  R.~Cooke, M.~Pettini, R.~A.~Jorgenson, M.~T.~Murphy and C.~C.~Steidel,
  Ap. J. {\bf 781}, 31 (2014)
  [arXiv:1308.3240 [astro-ph.CO]].

  \bibitem{riemer}
   S.~Riemer-S{\o}rensen {\it et al.},
  Mon.\ Not.\ Roy.\ Astron.\ Soc.\  {\bf 447}, 2925 (2015)
  [arXiv:1412.4043 [astro-ph.CO]].
  
  \bibitem{bala}
  S.~A.~Balashev, E.~O.~Zavarygin, A.~V.~Ivanchik, K.~N.~Telikova and D.~A.~Varshalovich,
  Mon.\ Not.\ Roy.\ Astron.\ Soc.\  {\bf 458}, no. 2, 2188 (2016)
  [arXiv:1511.01797 [astro-ph.GA]].
  
  \bibitem{cookeN}
  R.~J.~Cooke, M.~Pettini, K.~M.~Nollett and R.~Jorgenson,
  Astrophys.\ J.\  {\bf 830}, no. 2, 148 (2016)
  [arXiv:1607.03900 [astro-ph.CO]].
  
  \bibitem{riemer17}
  S.~Riemer-S{\o}rensen, S.~Kotu\v{s}, J.~K.~Webb, K.~Ali, V.~Dumont, M.~T.~Murphy and R.~F.~Carswell,
  Mon.\ Not.\ Roy.\ Astron.\ Soc.\  {\bf 468}, no. 3, 3239 (2017)
  [arXiv:1703.06656 [astro-ph.CO]].
  
  \bibitem{zava}
E. O. Zavarygin, J. K.  Webb, V. Dumont, S. Riemer-S{\o}rensen,
 Mon.\ Not.\ Roy.\ Astron.\ Soc.\  {\bf 477}, no. 4, 5536 (2018)
 [arXiv:1706.09512 [astro-ph.GA]].
 
 \bibitem{CPS}
  R.~J.~Cooke, M.~Pettini and C.~C.~Steidel,
  Astrophys.\ J.\  {\bf 855}, no. 2, 102 (2018)
  [arXiv:1710.11129 [astro-ph.CO]].

\bibitem{cfo5}
R.~H.~Cyburt, B.~D.~Fields and K.~A.~Olive,
JCAP \textbf{11}, 012 (2008)
[arXiv:0808.2818 [astro-ph]].

\bibitem{Fields:2022mpw}
B.~D.~Fields and K.~A.~Olive,
JCAP \textbf{10}, 078 (2022)
[arXiv:2204.03167 [astro-ph.GA]].

  \bibitem{ryan2000}
  S.~G.~Ryan, T.~C.~Beers, K.~A.~Olive, B.~D.~Fields and J.~E.~Norris,
  Astrophys.\ J.\  {\bf 530}, L57 (2000);
  L.~Sbordone, P.~Bonifacio, E.~Caffau, H.-G.~Ludwig, N.~T.~Behara, J.~I.~G.~Hernandez, M.~Steffen and R.~Cayrel {\it et al.},
  Astron.\ Astrophys.\  {\bf 522}, A26 (2010)
  [arXiv:1003.4510 [astro-ph.GA]].


\bibitem{Salvati:2015wxa}
L.~Salvati, L.~Pagano, R.~Consiglio and A.~Melchiorri,
JCAP \textbf{03}, 055 (2016)
[arXiv:1507.07243 [astro-ph.CO]].

\bibitem{CMB-S4:2016ple}
K.~N.~Abazajian \textit{et al.} [CMB-S4],
[arXiv:1610.02743 [astro-ph.CO]].

\bibitem{Wilson:2020lfr}
J.~T.~Wilson, D.~J.~Lawrence, P.~N.~Peplowski, V.~R.~Eke and J.~A.~Kegerreis,
Phys. Rev. C \textbf{104}, no.4, 045501 (2021)
[arXiv:2011.07061 [nucl-ex]].

\bibitem{Wietfeldt:2022tma}
F.~E.~Wietfeldt, R.~Biswas, J.~Caylor, B.~Crawford, M.~S.~Dewey, N.~Fomin, G.~L.~Greene, C.~C.~Haddock, S.~F.~Hoogerheide and H.~P.~Mumm, \textit{et al.}
[arXiv:2209.15049 [nucl-ex]];
see also: \href{https://npl.illinois.edu/research/UCNtau}{https://npl.illinois.edu/research/UCNtau}


\end{thebibliography}
\end{document}